\documentclass[journal]{IEEEtran}
\usepackage[section] {placeins}
\usepackage{graphicx}
\usepackage{amsmath}
\usepackage{subfigure}
\usepackage{rotating}
\usepackage{multirow}
\usepackage{array}
\usepackage{rotating}
\usepackage{auto-pst-pdf}
\usepackage{graphicx}

\begin{document}

\title{DREEM-ME: Distributed Regional Energy Efficient Multi-hop Routing Protocol based on Maximum Energy in WSNs}

\author{N. Amjad$^{1}$, N. Javaid$^{1,2}$, A. Haider$^{1}$, A. A. Awan$^{1}$, M. Rahman$^{3}$\\\vspace{0.4cm}
$^{1}$Dept of Electrical Engineering, COMSATS Institute of IT, Islamabad, Pakistan.\\
$^{2}$CAST, COMSATS Institute of IT, Islamabad, Pakistan.\\
$^{3}$Abasyn University, Peshawar, Pakistan.}
% make the title area
\maketitle

\begin{abstract}
\boldmath
Wireless distributed sensor network consists of randomly deployed sensors having low energy assets. These networks can be used for monitoring a variety of environments. Major problems of these networks are energy constraints and their finite lifetimes. To overcome these problems different routing protocols and clustering techniques are introduced. We propose DREEM-ME which uses a unique technique for clustering to overcome these two problems efficiently. DREEM-ME elects a fix number of cluster heads (CHs) in each round instead of probabilistic selection of CHs. Packet Drop Technique is also implemented in our protocol to make it more comprehensive and practical. In DREEM-ME confidence interval is also shown in each graph which helps in visualising the maximum deviation from original course. Our simulations and results show that DREEM-ME is much better than existing protocols of the same nature.

\end{abstract}
\IEEEpeerreviewmaketitle

\section{Introduction}
\IEEEPARstart{A}{Wireless}
 Sensor Network(WSN) is a collection of small randomly dispersed devices that provide the ability to monitor physical and environmental conditions in real time, such as, temperature, pressure, light and humidity and the ability to provide efficient and reliable communications via a wireless link. WSNs are undergoing a revolution that promises to have a significant impact throughout society. These networks consist of battery powered nodes that are endowed with a multitude of sensing modalities. WSNs are used in area monitoring, landslide detection, agriculture, security, medical applications and environmental monitoring.
 WSNs are independent when deployed into the field because they have the ability of self-configuration and survival. Once they are deployed in the field they organize themselves in the shape according to the applied protocol and then start communicating with each other. As nodes have limited amount of initial energy, they are bound to operate within such constraints. Then CHs are elected and they collect data from their corresponding child nodes, aggregate it and then send it to Base Station (BS). The two challenges for routing protocols are:

\begin{itemize}
  \item To select the best route to send the data to the BS or the CH.
  \item CH selection technique which is used in the network.
\end{itemize}

Nowadays, research challenge in WSNs is to deal with low power communications. Efficient energy utilization requires the protocol to be more systematic which can select the best possible way to send the aggregated data to the BS, so, the energy consumption is minimum. Old fashioned routing techniques like Direct Communication, MTE, LEACH and LEACH-C are not as much efficient as present protocols are. There are two types of clustering:
\begin{itemize}
  \item Static Clustering
  \item Dynamic Clustering
\end{itemize}
 Clusters once constructed and never be changed throughout network lifetime, are called Static Clusters, while clusters based on some sort of network characteristics and are changing during network operation are known as Dynamic Clusters [10].

DREEM-ME is based upon static clustering in which maximum energy based CH selection is used. Network is clearly divided into concentric circles. Total regions are 9 which are represented by Region 1 to Region 9 as shown in Figure 1. All the regions make CHs of their own and start communication except the Region 1 in which nodes directly communicate with the BS. In our network:
\begin{itemize}
  \item All the nodes are homogeneous (Initially having same energy).
  \item All nodes are proactive (continually monitoring data).
  \item BS is in the centre of the network.
\end{itemize}
In real world applications of WSNs, the wireless communication between nodes and BS does not behave ideally. Packet loss occurs when one or more packets traveling across a wireless link fails to reach their destination. This is because some of the packets are lost due to some risk factors like interference, attenuation, noise, etc. The wireless link can sometimes be in the bad state and as a result some of the packets may not be received at the destination. So, we have discussed and implemented the Random Uniformed Model[2] of packet loss in DREEM-ME to make it more close to reality and more practical.

%The remainder of this paper is organized as follows. In Section II, Motivation for this protocol is given. Section III contains a brief introduction of the radio model. Section IV shows our proposed scheme in details. Section V elaborates all the results and comparison of DREEM-ME. Section VI addresses the future work. Finally, Section VII concludes our research.

\section{Motivation}
Every node in the WSNs is crucial and has its own importance. Each node possess a little amount of energy and that is not rechargeable, so, energy must be used efficiently for the sake of network lifetime. But Previous works on WSNs such as LEACH [1], TEEN [13], SEP [14], DEEC [15] and LEACH-C [16] show the possibility of coverage holes during lifetime of the network and that is unacceptable. Clustering technique of LEACH [1] does not assure a fix number of CHs in each round therefore, its behavior is not so appreciable in case of network lifetime. LEACH protocol have selected the CHs on the basis of probability, so, the number of CHs selected are not optimum. This leads to inefficient use of energy. In our proposed technique, we select the nodes as CHs which carry maximum energy in a particular region. So, this technique assures the optimum number of cluster heads in every round. Also we have used maximum energy node as the CH of every region this also increases efficiency of the system. In HEER [18], CH selection is based on the ratio of residual energy of node and average energy of network.

For efficient use of energy and improvement of coverage, the protocol divides the total area into small sub-regions and these sub-regions are treated separately for the nodes distribution and it helps improving the coverage of the network. Nodes are not deployed in any pattern [3] but are deployed randomly in every region. In this way, the CHs will also be optimum in every round. As shown by the research that 8 cluster heads are optimum in the network [12]. And other problems with previous protocols are their stability period and network lifetime, which are not good enough. So, DREEM-ME is proposed as its lifetime and stable region are far better than previous protocols.

\section{First Order Radio Model}
 In our work, we assume a simple first order radio model in which the radio dissipates $E_{elec}$ = 50 nJ/bit for powering the transmitter or receiver circuitry and $\varepsilon_{amp}$ = 100 pJ/bit/m2 for the transmit amplifier to achieve an acceptable Eb/No. We also take in account the $d^2$ energy loss due to channel transmission. Thus, to transmit a k-bit message a distance d using the radio model will consume:
\begin{equation}
  E_{Tx}(k,d)=E_{Tx-Elec}(k)+E_{Tx-amp}(k,d)
\end{equation}
\begin{equation}
  E_{Tx}(k,d)=E_{elec} * k + E_{amp}*k*d^2
\end{equation}
 And for receiving a k-bit message will consume:
\begin{equation}
  E_{Rx}(k,d)=E_{Rx-Elec}(k)
\end{equation}
\begin{equation}
  E_{Rx}(k)=E_{elec} * k
\end{equation}

This model shows that for the same SNR the energy consumed in transmitting a k-bit message from node 1 to node 2 is the same as the energy consumed in transmitting same message from node 2 to node 1. Transmitter circuitry of the CH node also consumes $E_{DA}$ energy equivalent to aggregate the received data from its child nodes. Different transmission schemes are discussed in [9].

\section{Proposed Scheme : DREEM-ME}
There exists a trade off between coverage and the energy. The target which we want to achieve is to maximize the network coverage. So, we have localized the whole network and divided the network into sub-regions that helps in avoiding the coverage hole [8][11]. The following are the main parts of our proposed model:

\subsection{Network Model}

In our Model, we have taken 100mx100m area for the wireless sensor network and then divided it by three concentric circles with the center at origin [4][6]. The BS is placed at the [0,0] coordinates. The radius of the innermost, middle and outer circles are 20m, 35m and 50m. These circles are again divided in sectors to make optimal regions for our network. Then the next step is to make sectors of the outer two circles by 90 degrees. So, 9 regions are formed in this way and then we deployed all 90 nodes in our area. The nodes are divided equally in 9 regions , so, every region gets 10 nodes fixed in every round. The 10 nodes of every region will be deployed in their corresponding regions randomly. Dividing the network into subregions will help in reducing the distance between cluster members and CHs.

\subsection{ Clustering And Routing Techniques }

The previous protocols are using the probabilistic techniques for CH selection [1][5]. Nodes then associate with each CH based upon received signal strength. However, in DREEM-ME the CH selection is entirely based upon the maximum energy. In a particular region, the node with maximum energy is selected as the CH for that region in the current round. So, in this way the burden of aggregating the data of 9 nodes and then sending it to the next CH or the BS, is handled by the node with maximum energy.

Association is also important in energy utilization because if any node is forced to send its data to the cluster head of its own region while that cluster head is at a greater distance to that particular node than any other cluster head then it is not efficient for energy of the network. So, in DREEM-ME a unique technique for association of nodes is applied. All the nodes of outer regions (i.e. 6, 7, 8, 9) of our network which are not CHs check their distances from the CHs of six regions which are close to them. For example, each node of Region 6 checks its distance from CHs of its nearby regions i.e. 2, 3, 5, 6, 7, 9 and then finds the minimum of these six distances. In this way every node of region 6 sends its data to the CH which is at minimum distance. This leads to the increase in energy efficiency which is our main goal [7][9][17].

Routing is the backbone of the protocol because the consumption of energy depends upon routing [19]. In DREEM-ME, 10 nodes of the Region 1 are using Direct communication because they are at smaller distance to the BS as compared to the nodes in other sub-regions. And all the other sub-regions are considered as static clusters of 10 nodes each.

 As, the CHs of the outermost circles are at a long distance from the BS in the case of direct communication, so, in order to make it more energy efficient, multi-hop technique is used in our proposed protocol. Nodes of regions 2 to 9 select their CHs by checking the maximum energy, those CHs then collect the data from their child nodes and aggregate it. Here, CHs of Regions 9 to 6 send their aggregated data to the CHs of the Regions 5 to 2. CHs of Regions 2 to 5 receive the data packets of all their child nodes and also from CHs of the Regions 6 to 9 respectively. Finally, CHs of the regions 2 to 5 send their aggregated data to the BS.

\begin{figure*}[t]
\centering
\includegraphics[height=12cm , width=11cm]{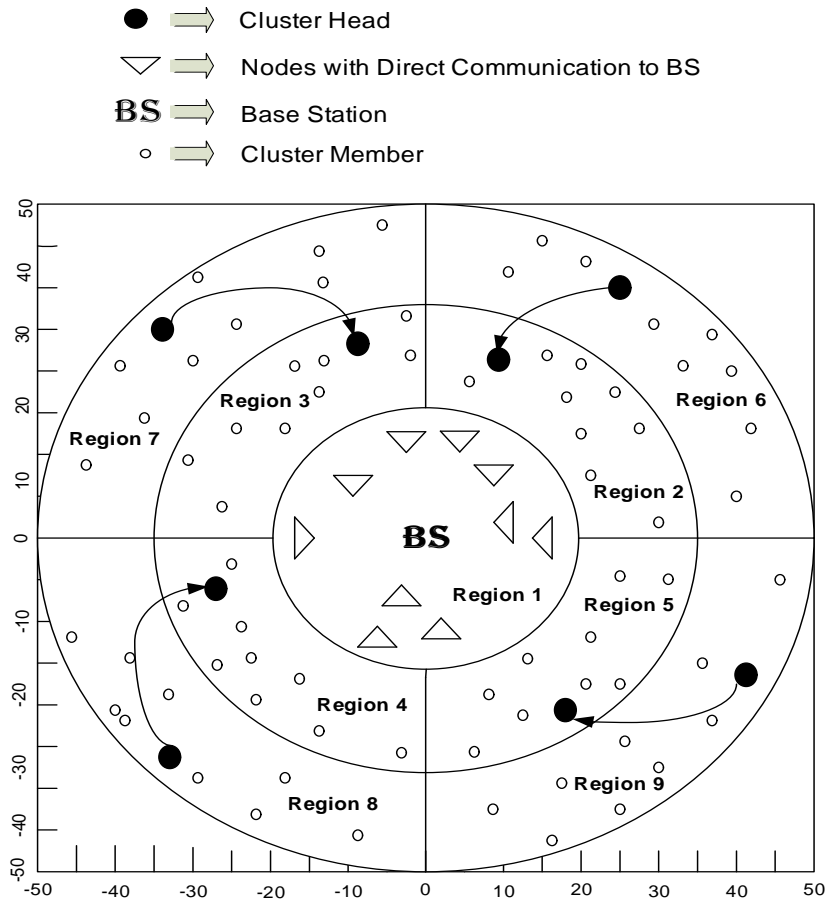}
\caption{Division of Area}
\end{figure*}

\section{Simulations and Results}
 In this section we are going to discuss the efficiency and performance of our purposed protocol. We take a 100m x 100m area for our network and total nodes are 90. We divided the area into three concentric circles with 20m, 35m and 50m radii and then made regions out of these three circles as shown in the Figure 1. We provided 10 nodes to each region of our network. Some basic simulation parameters are given in Table 1.

\begin{table}[ht]
\caption{\bf Parameters used in Simulations} % title of Table
\centering % used for centering table
\begin{tabular}{|c| c|} % centered columns (2 columns)
\hline
Parameter & Value \\ [0.5ex] % inserts table
%heading
\hline % inserts single horizontal line
Network size & 100m x 100m \\ % inserting body of the table
Total Nodes & 90 \\
Initial energy of each node & 0.5J \\
$E_{TX}$ & 50nJ \\
$E_{RX}$ & 50nJ \\
$E_{DA}$ & 5nJ \\
Maximum Radius of Circles & 50 \\
Packet size & 4000 bits \\ [1ex] % [1ex] adds vertical space
\hline %inserts single line
\end{tabular}
\end{table}

\subsection{ Simulation Parameters }

Parameters of simulations are discussed below:

\subsubsection{ Random Uniformed Packet Drop }
In DREEM-ME, we implemented random uniformed model of packet loss. This model specifies the dropped packets during the transmission. Normally, researchers do not consider the fact that the wireless links are not ideal therefore, they are not capable of sending 100 \% of the data successfully to the BS due to a lot of risk factors affecting the transmission. These factors include interference, noise, attenuation and reflection etc. Aside from simulations, the real world applications of WSNs, the wireless links between nodes and BS does not behave ideally. Packet loss occurs when one or more packets of data traveling across a wireless link fails to reach their destination. The wireless link can sometimes be in the bad state and as a result maybe some packets may not be received at the destination. So, we have implemented the Random Uniformed Model [2] of packet loss in DREEM-ME to make it more close to reality and more practical. In DREEM-ME the probability of packet loss is 0.3 and packet delivery in 0.7 as shown in Figure 5. So, simulations of DREEM-ME, LEACH and LEACH-C are taken 5 times and then calculated the average of dropped packets in each round of their lifetime.

\subsubsection{ Confidence Interval }
In WSNs the deployment of the nodes is random and in every round of the simulation, nodes are placed on different locations in the network area, so, as a result the energy consumption of each node varies in every round. So, each time we simulate the results are different. These results fluctuate to-and-fro around a mean value in every simulation. So, taking this thing into account we have simulated our protocol 5 times, averaged the value, calculated confidence interval of Dead nodes, Alive nodes, Packets sent to BS per round and Packets received at BS and Dropped packets per round, then plotted all of them in Figures 2, 3, 4, 5. In statistics, a confidence interval is a type of interval estimate of data and is used to indicate the reliability of an estimate. Confidence interval is the interval in which we are pretty confident about our results simply. It is calculated from the observations that frequently includes the parameter of interest if the experiment is repeated. More specifically, the meaning of the term "Confidence Interval" is that, if confidence intervals are constructed across many separate data analysis of repeated experiments, the proportion of such intervals that contain the true value of the parameter will match the confidence interval; this is guaranteed by the reasoning underlying the construction of confidence interval. Whereas two-sided confidence limits form a confidence interval, their one-sided counterparts are referred to as lower or upper confidence bounds confidence interval. Confidence interval consist of a range of values that act as good estimates of our values of interest. So, we have observed the range of variance of our desired results and then defined their upper and lower values and the mean also so that we can plot their confidence intervals.

\subsubsection{ Dead Nodes }

All Nodes remain alive until their energy is greater than zero. LEACH [1] uses its own probability function for clustering in the whole area and all nodes possess the same probability to become a CH therefore, all nodes die linearly after the first node dies. Whereas, DREEM-ME is using direct communication to the BS in its 1st region and all other regions use clustering which is based upon maximum energy. It means the node which has maximum energy in its corresponding region is selected as the CH, this technique ensures the energy efficiency of the system. First of all nodes of regions of outermost circle die and after that nodes of regions of middle circle die and the Direct Communication nodes die in the last because they are much closer to the BS. Results show that the stability region of DREEM-ME is 40\% better than LEACH.

\begin{figure}[!h]
\centering
\includegraphics[height=7cm, width=9cm]{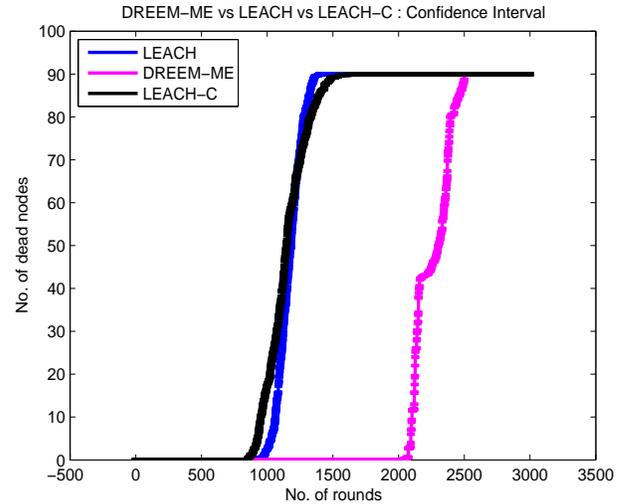}
\caption{ Dead Nodes }
\end{figure}

\subsubsection{ Packets Sent To BS }
The Figure 3 shows five times averaged values of the total number of packets sent to BS per round of the network lifetime of LEACH, LEACH-C and DREEM-ME. According to our network strategy packets sent to the BS per round should ideally follow the explanation below:
      Packets sent to BS by 1st Region DT nodes = 10
      Packets Sent to BS by 2nd Region CH Node = 1
      Packets Sent to BS by 3rd Region CH Node = 1
      Packets Sent to BS by 4th Region CH Node = 1
      Packets Sent to BS by 5th Region CH Node = 1
      Total Packets Sent to BS per round = 14
	So, as long as all the nodes are alive the packets sent remain 14. When nodes of outer two circles start to die the number of packets gradually decrease till 2154th round and at that time only direct communication nodes are left. Now graph is stable until 2370th round and after that Region 1 nodes start to die the number of packets start to decrease. Whereas, LEACH is using clustering in its network area of 100mx100m and every node has the same probability to become a CH. LEACH does not assures how many CHs will be formed during any round and in every round the number of CHs are varying around 9 (p=0.1) when all nodes are alive, and packets start to decrease as soon as first node dies at 855th round. So, the LEACH protocol forms approximately 9 CHs in its every round, so, the packets sent to BS should also be 9. So, as shown by the Figure 3 the Packets sent by LEACH in every round should be around 9 until first node dies.

\begin{figure}[!h]
\centering
\includegraphics[height=7cm, width=9cm]{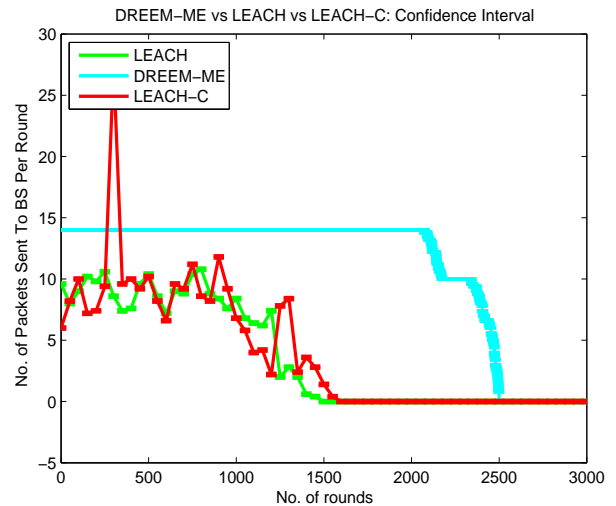}
\caption{ Packets Sent to Base Station }
\end{figure}

\subsubsection{Packets Received By BS }
In DREEM-ME Packet Drop concept is used which makes it more close to the reality situation. Because in reality the wireless links are not perfect or ideal, therefore, there is always a probability that some of packets may be dropped on their way. So, the graph below shows that packets received are not the same as the packets sent in the same round. As, the nodes start to die the packets received also decrease. LEACH makes variable CHs, so, its received packets are also varying. DREEM-ME has 14 nodes that send their data packets to BS therefore, this graph of DREEM-ME shows that maximum packets received are 14. We have used Uniform Random Model of packet drops that says there is a probability that a packet is dropped during the transmission and may not be received by the BS. Confidence interval is also introduced in the received packets by BS that shows the averaged range of possible values of the received packets because in every simulation the results are different, so, this makes it more general and authenticated result. There are also some peaks shown by the graph of LEACH, it is because LEACH does not assure the maximum number of CHs in the network therefore, in some rounds packets received are more than expected because of the CHs variance.

In Figure 4 the number of received packets by the BS are shown. DREEM-ME sends 14 packets in each round and the probability of packet drop is 0.3, so, received packets by BS are fluctuating around 10 because almost 30 \% of packets are dropped during flight. Whereas, LEACH [1] sends 10 packets in each round and the packet drop probability is 0.3, so, the received packets are varying around 7 in the stability region. The anonymous peaks in the LEACH graph are because LEACH possess a unique behavior of clustering and allows more than 9 CHs in some rounds hence the peaks come.

\begin{figure}[!h]
\centering
\includegraphics[height=7cm, width=9cm]{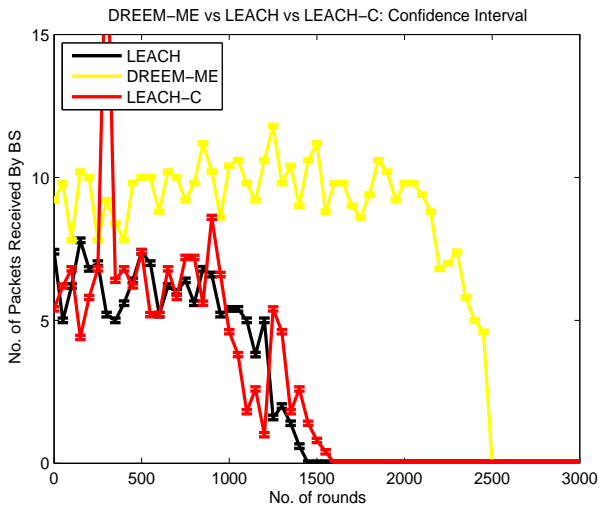}
\caption{ Packets Received By Base Station }
\end{figure}

\subsubsection{ Packets Dropped }

The number of packets dropped on the link in every round of both LEACH and DREEM-ME are shown in Figure 5. We calculated the dropped packets with the probability of dropping as 0.3 out of 1 but it is also possible practically that the probability of packet loss is less than 0.3.

\section{Future Work}
We are working on some more clustering and routing techniques to make the network much better and more efficient than DREEM-ME. In future we would like to reduce deficiencies which are expected in this paper and implementation of DREEM-ME in other clustering protocols like Threshold sensitive energy efficient sensor network protocol [13], stable election protocol [14], distributed energy efficient clustering [15], etc. In future, we will try to make our network heterogeneous as done in [14][15][20] for much better performance.

\begin{figure}[!h]
\centering
\includegraphics[height=7cm, width=9cm]{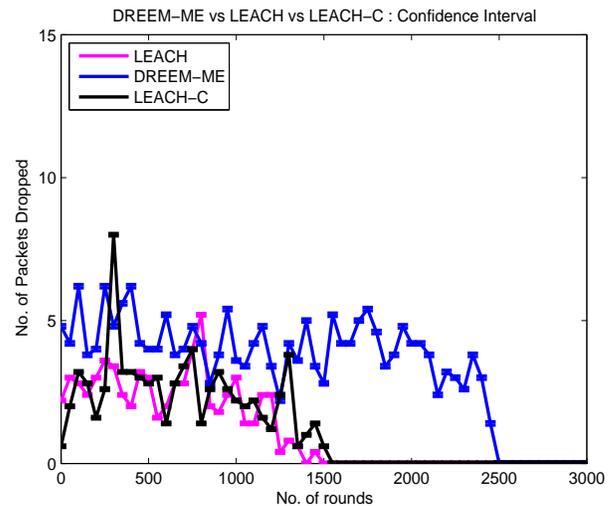}
\caption{ Packets Dropped }
\end{figure}

\section{Conclusion}
In this paper, we have proposed a new clustering technique for WSNs. DREEM-ME uses static clustering and maximum energy based CH selection. Multi-hop route is used for the CHs at long distance to sink. Good thing about DREEM-ME is the network field is divided evenly into circles and sectors to reduce the distance between CHs and BS. In MATLAB simulations we compared our results with LEACH and LEACH-C. In terms of achieving optimum number of CHs in each round and CH selection technique of our technique provided better results than its counterparts, in terms of network lifetime, stability period, area coverage and throughput.

\end{document}